\documentclass[aps,prx,reprint,superscriptaddress,showkeys]{revtex4-1}

\usepackage{amsmath}
\usepackage{graphicx}
\usepackage[colorlinks=true, linkcolor=blue, citecolor=blue, urlcolor=magenta]{hyperref}

\begin{document}

\title{Solar dynamo as host power pacemaker of the Earth global climate}

\date{30 June 2011}

\author{V.D. Rusov}
\email{Corresponding author. E-mail: siiis@te.net.ua}
\affiliation{Department of Theoretical and Experimental Nuclear Physics, Odessa National Polytechnic University, Odessa, Ukraine}

\author{E.P. Linnik}
\affiliation{Department of Theoretical and Experimental Nuclear Physics, Odessa National Polytechnic University, Odessa, Ukraine}

\author{V.N. Vaschenko}
\affiliation{State Ecological Academy for Postgraduate Education and Management, Kiev, Ukraine}

\author{S. Cht. Mavrodiev}
\affiliation{Institute for Nuclear Research and Nuclear Energy, BAS, Sofia, Bulgaria}

\author{M.E. Beglaryan}
\affiliation{Department of Theoretical and Experimental Nuclear Physics, Odessa National Polytechnic University, Odessa, Ukraine}

\author{T.N. Zelentsova}
\affiliation{Department of Theoretical and Experimental Nuclear Physics, Odessa National Polytechnic University, Odessa, Ukraine}

\author{V.A. Tarasov}
\affiliation{Department of Theoretical and Experimental Nuclear Physics, Odessa National Polytechnic University, Odessa, Ukraine}

\author{D.A. Litvinov}
\affiliation{Department of Theoretical and Experimental Nuclear Physics, Odessa National Polytechnic University, Odessa, Ukraine}

\author{V.P. Smolyar}
\affiliation{Department of Theoretical and Experimental Nuclear Physics, Odessa National Polytechnic University, Odessa, Ukraine}

\author{B. Vachev}
\affiliation{Institute for Nuclear Research and Nuclear Energy, BAS, Sofia, Bulgaria}

\begin{abstract}
It is known that the so-called problem of solar power pacemaker related to possible existence of some hidden but key mechanism of energy influence of the Sun on fundamental geophysical processes is one of the principal and puzzling problems of modern climatology. The "tracks" of this mechanism have been shown up in different problems of solar-terrestrial physics for a long time and, in particular, in climatology, where the solar-climate variability is stably observed. However, the mechanisms by which small changes in the Sun's energy (solar irradiance or insolation) output during the solar cycle can cause change in the weather and climate are still unknown. 

We analyze possible causes of the solar-climate variability concentrating one's attention on the physical substantiation of strong correlation between the temporal variations of magnetic flux of the solar tachocline zone and the Earth magnetic field (Y-component). We propose an effective mechanism of solar dynamo-geodynamo connection which plays the role of the solar power pacemaker of the Earth global climate.
\end{abstract}


\maketitle

\section{Introduction}

\begin{figure*}
  \includegraphics{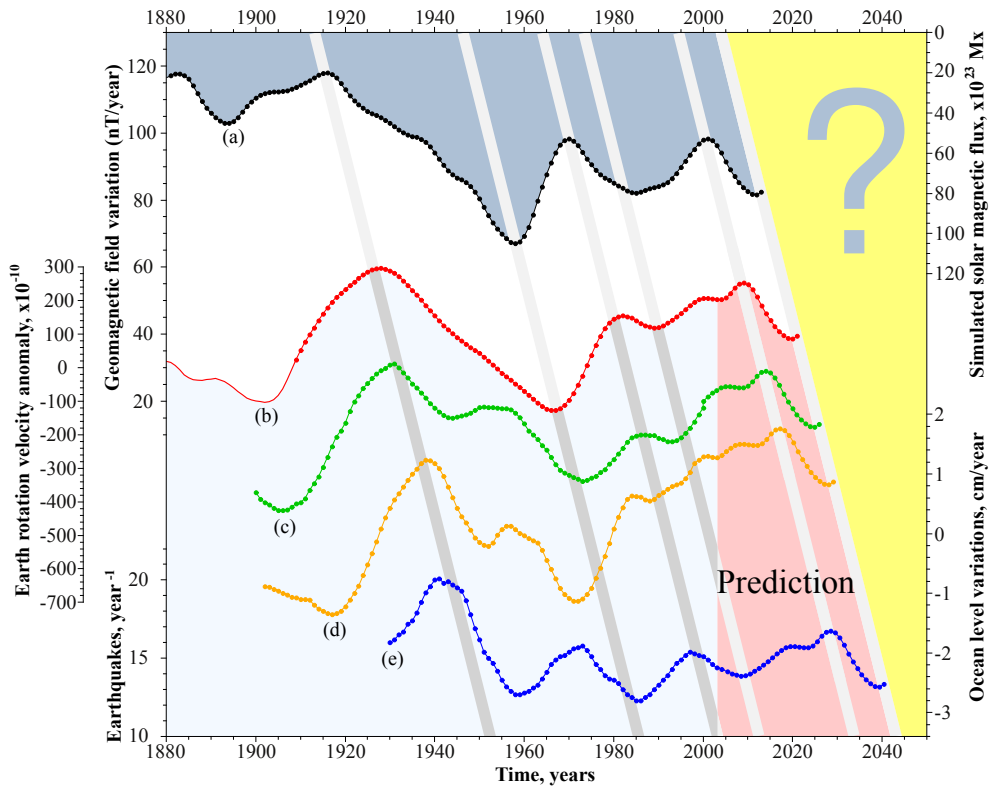}
  \caption{\label{fig1}Time evolution (a) the variations of magnetic flux in the bottom (tachocline zone) of the Sun convective zone (see Figure 7f \cite{ref3}), (b) of the geomagnetic field secular variations (Y-component, nT/year), whose values are obtained at the Eskdalemuir observatory (England) \citep{ref4}, where the variations $(\delta Y / \delta t)$ are directly proportional to the westward drift of magnetic features, (c) the variation of the Earth's rotation velocity \cite{ref5}, (d) the variations of the average global ocean level (PDO+AMO, cm/year) \cite{ref6} and (e) the number of large earthquakes (with the magnitude $M \geq 7$) \cite{ref7}. All curves are smoothed by sliding intervals in 5 and 11 years. The pink area is the prediction region. Note: formation of the second peaks on curves (c)-(e) is mainly predetermined by nuclear tests in 1945-1990.}
\end{figure*}

It is known that in spite of a long history the nature of the energy source maintaining a convection in the liquid core of the Earth or, more exactly, the mechanism of the magnetohydrodynamic dynamo (MHD) generating the magnetic field of the Earth still has no clear and unambiguous physical interpretation (see \cite{ref1} and refs. therein). The problem is aggravated because of the fact that none of candidates for an energy source of the Earth magnetic-field \cite{ref1} (secular cooling due to the heat transfer from the core to the mantle, internal heating by radiogenic isotopes, e.g., 40K, latent heat due to the inner core solidification, compositional buoyancy due to the ejection of light element at the inner core surface) can't in principle explain one of the most remarkable and mystic phenomena in solar-terrestrial physics, which consists in strong (negative) correlation \footnote{Note that the strong (negative) correlation between the temporal variations of magnetic flux in the tachocline zone and the Earth magnetic field (Y-component) will be observed only for experimental data obtained at that observatories where the temporal variations of declination ($\delta D / \delta t$) or the closely associated east component ($\delta Y / \delta t$) are directly proportional to the westward drift of magnetic features \cite{ref2}. This condition is very important for understanding of physical nature of indicated above correlation, so far as it is known that just motions of the top layers of the Earth's core are responsible for most magnetic variations and, in particular, for the westward drift of magnetic features seen on the Earth's surface on the decade time scale. Europe and Australia are geographical places, where this condition is fulfilled (see Figure 2 in \cite{ref2}).} \cite{ref2} between temporal variations of the magnetic flux in the tachocline zone (the bottom of the Sun convective zone) \cite{ref3} and the Earth magnetic field \cite{ref4} (Figure \ref{fig1}).

At the same time, supposing that the transversal (radial) surface area of tachocline zone, through which a magnetic flux passes, is constant in the first approximation, we can consider that the magnetic flux variations describe also the magnetic field temporal variations in the tachocline zone of the Sun. In this sense, it is obvious that a future candidate for an energy source of the Earth magnetic field must play not only the role of a natural trigger of solar-terrestrial connection, but also directly generate the solar-terrestrial magnetic correlation by its own participation.
 
The fact that the solar-terrestrial magnetic correlation has, undoubtedly, fundamental importance for evolution of all the Earth's geospheres is confirmed by existence of a stable and strong correlation between the temporal variations of the Earth magnetic field, the Earth angular velocity, the average global ocean level and the number of large earthquakes (with the magnitude M$\geq$7), whose generation is apparently predetermined by a common physical cause of unknown nature (see Figure \ref{fig1}).

On the other hand, it is clear that understanding of the mechanism of solar-terrestrial magnetic correlation can become the clue of so-called problem of solar power pacemaker related to possible existence of some hidden but key mechanism of energy influence of the Sun on the fundamental geophysical processes. It is interesting, that the "tracks" of this mechanism have been observed for a long time and manifest themselves in different problems of solar-terrestrial physics and, in particular, in climatology, where the mechanisms by which small changes in the Sun's energy output during the solar cycle can cause change in the weather and climate have been a puzzle and the subject of intense research in recent decades.
Thus it becomes obvious that purposeful or unpurposed neglect of the mechanism of solar power pacemaker in any point or multi-zonal models of the Earth global climate can result in serious errors in interpretation of the experimental temperature and other geophysical trends, especially in the compilation of short-term and, all the more, long-term forecasts.

In this paper we consider hypothetical particles ($^{57}$Fe solar axions) as the main carriers of the solar-terrestrial connection, which can transform into photons in external fluctuating electric or magnetic fields by virtue of the inverse coherent Primakoff effect \citep{ref8}. At the same time we ground and develop the axion mechanism of solar dynamo -- geodynamo connection, where the energy of axions is modulated at first by the magnetic field of the solar tachocline zone (due to the inverse coherent Primakoff effect), and after that is resonantly absorbed in the iron core of the Earth, thereby playing the role of an energy source and modulator of the Earth magnetic field. Justification of the axion mechanism of solar dynamo -- geodynamo connection and its account within the framework of the bifurcation model of the Earth global climate on different time scales is the goal of this article.

\section{Peculiarities of the bifurcation model of the Earth global climate on different time scales}

As is shown in our papers \cite{ref9,ref10}, the basic equation of energy-balance model of the Earth global climate is the bifurcation equation (with respect to the Earth surface temperature (see Figure \ref{fig2})) of assembly-type catastrophe with two governing parameters, which describe insolation variations and the Earth magnetic field variations (or the variations of cosmic ray intensity in the atmosphere). A general bifurcation problem of this energy-balance model (see equations (20)-(23) and (26)-(28) in \cite{ref9,ref10}), which consists in determination of the global temperature $T(t)$ and its increment $\Delta T(t)$, is reduced to finding the stable solution set of equations:

\begin{figure}
  \includegraphics{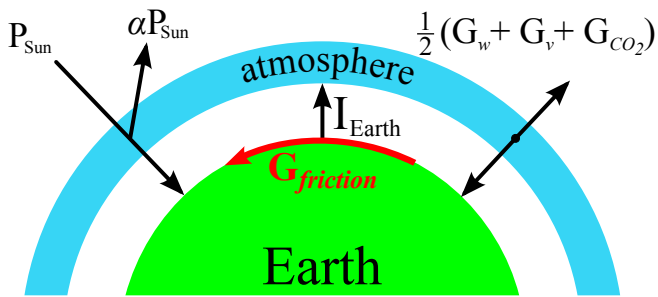}
  \caption{\label{fig2}The energy fluxes balance on the Earth surface. Here $G_w$, $G_v$, $G_{CO}$ are the heat energy power re-emitted 2 by liquid water, water vapour and carbon dioxide, respectively, $G_{friction}$ is the heat or dissipation energy generated by the Earth surface-to-atmosphere bottom layer friction.}
\end{figure}

\begin{equation}
   \label{eq1}
    \frac{\partial }{\partial T} U^{*}(T,t) = T^{3}_{t} + a(t) \cdot T_{t} + b(t) = 0,
\end{equation}

\noindent where

\begin{equation}
   \label{eq2}
    a(t) = - \frac{1}{4 \delta \sigma}a_{\mu} H_{\oplus}(t),
\end{equation}

\begin{equation}
   \label{eq3}
    b(t) = - \frac{1}{4 \delta \sigma} \left[ \frac{\eta_{\alpha} S_0}{4} + \frac{1}{2} \beta + \frac{1}{2} b_{\mu} H_{\oplus} (t) \right]
\end{equation}

\noindent and

\begin{equation}
   \label{eq4}
   \frac{\partial}{\partial T} \Delta U^{*}(\Delta T,t) \cong \Delta T^{3}_{t} + \tilde{a}(t) \cdot \Delta T_{t} + \tilde{b}(t) = 0,
\end{equation}

\noindent where
	
\begin{equation}
   \label{eq5}
   \tilde{a}(t) = - \frac{37.6}{\sigma T_t^3} a_{\mu} H_{\oplus}(t) = - \tilde{a}_0 H_{\oplus}(t);
\end{equation}

\begin{widetext}
  \begin{equation}
    \label{eq6}
      \begin{matrix}  
    \tilde{b}(t) & = - \dfrac{37.6}{\sigma T_t^3} \left[ \eta_{\alpha} \dfrac{S_0 + \Delta \hat{W}(t) \sigma_s}{4} - 4\delta \sigma T_t^3 + \dfrac{1}{2} \beta + \dfrac{1}{2} \left( 2a_{\mu} T_t + b_{\mu} \right) H_{\oplus}(t) \right] = \\
  & - \tilde{b}_0 \left[ \eta_{\alpha} W_{reduced}(t) - 4 \delta \sigma T_t^3 + \dfrac{1}{2} \beta + \dfrac{1}{2} \left( 2a_{\mu} T_t + b_{\mu} \right) H_{\oplus}(t) \right];
      \end{matrix}
  \end{equation}
\end{widetext}

$U^*(T, t)$ describes with an accuracy up to constant the so-called "inertial" power of heat variations in the Earth climate system; $\Delta U^*(T, t)$ is the variation of $U^*(T, t)$; $H_{\oplus}$ is the relative intensity of terrestrial magnetism, $T_t$ is the average global temperature of the Earth surface at the time t, K; $\Delta T_t$ is the variation of $T_t$; $S_0 = 1366.2~Wm^{-2}$ is "solar constant"; $\delta$ = 0,95 is coefficient of gray chromaticity of the Earth surface radiation; $\sigma = 5,67 \cdot 10^{-8}$ is the Stephen-Boltzmann constant, $Wm^{-2}K^{-4}$; $\eta_{\alpha} = 0,0295 K^{-1}$; $\beta$ is the accumulation rate of carbon dioxide in the atmosphere normalized by unit of temperature, $kgK^{-1}$; $a_{\mu}$ and $b_{\mu}$ are constants, whose dimensions are $Wm^{-2}K^{-2}$ and $Wm^{-2}K^{-1}$, respetively; $\Delta \hat{W}(t)$ is the insolation reduced normalized variation; $\sigma_s$ is the root-mean-square deviation; $W_{reduced} = S_0 + \Delta \hat{W}(t) \sigma_s$ is the reduced annual insolation. 

\begin{figure}
  \includegraphics{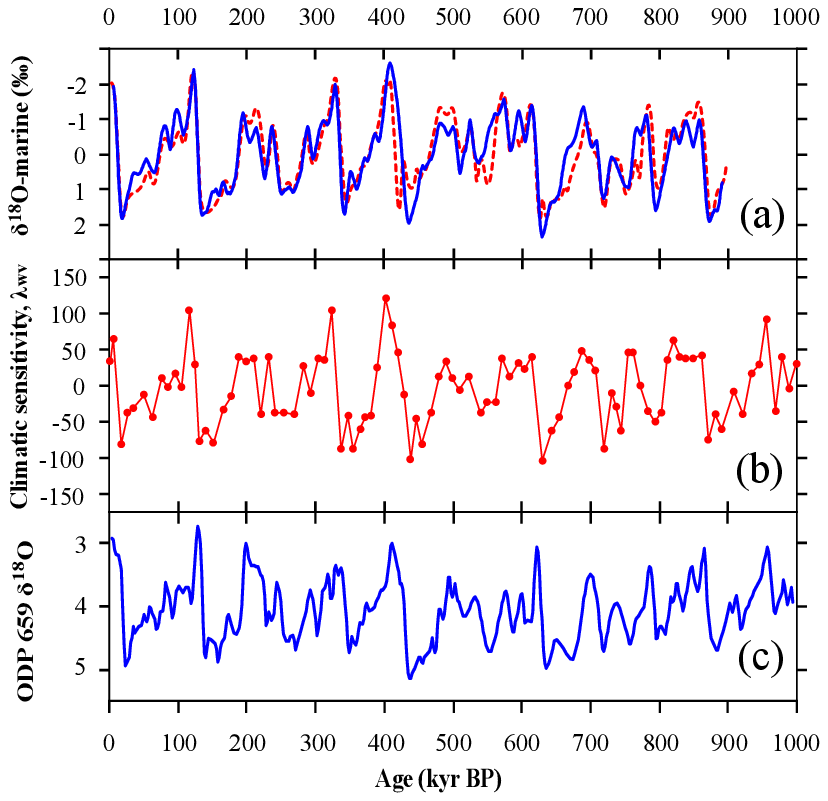}
  \caption{\label{fig3}Comparison of the theoretical time series of climatic sensitivity $\lambda_{w+v}$ calculated by equations (3) and (19) from \cite{ref10} (b) with the time series of $\delta^{18} O$ isotopic concentration (the conditional analogue of ice volume) measured in the deep-water experiments: (a) Bassinot et al. \cite{ref12} (solid blue line) and Imbrie et al. \cite{ref13} (dashed red line), (c) Tidemann et al. \cite{ref14}}
\end{figure}

Within the framework of proposed bifurcation model (i) comparison of the solution of energy-balance model of the Earth global climate and the EPICA Dome C and Vostok experimental data of the Earth surface palaeotemperature evolution over past 420 and 740 kyr is given; (ii) possible sharp warmings of the Dansgaard-Oeschger type during the last glacial period due to stochastic resonance is theoretically argued; (iii) the concept of climatic sensitivity of water in the atmosphere, whose temperature instability has the form of so-called hysteresis loop, is proposed, and based on this concept the time series of total fresh water mass (or vice versa the global ice volume) over the past 1000 kyr, which is in good agreement with the time series of $\delta^{18}$O concentration in sea sediments (Figure \ref{fig3}), is obtained; (iiii) groundlessness of the so-called "CO$_2$ doubling" problem is discussed. 

One of the main features of bifurcation model of the Earth global climate lies in the wonderful fact that priori knowledge of only two governing parameters, which are set by the known time series of insolation variations and variations of the Earth magnetic field (or the variations of cosmic ray intensity in the atmosphere) is required to solve the basic equations (\ref{eq1})-(\ref{eq6}) of energy-balance model (or to determine theoretical temperature trends). Other not less interesting feature of this model is the so-called principle of structural invariance, which means that the shape of global climatic potential (assembly-type catastrophe)

\begin{equation}
  \label{eq7}
  U(T,t) = \dfrac{1}{4} T^4 + \dfrac{1}{2} a(t) T^2 + b(t),
\end{equation}

is structurally invariant on different time scales. 
In other words, the principle of structural invariance of the balance equations of climate models evolving on the different time scales is not only a direct indicator of the correctly guessed physics of non-uniformly scaled processes, but it simultaneously specifies unambiguous rules for transition from one time scale to the other within the framework of one global model as well as for transition from the (one-zonal!) model of the Earth global climate on any time scale to the multizonal model of global climate or weather on a short time scale. It means that the system of equations of the multizonal model of global climate or weather convoluted into the balance equation of one-zonal model must fully keep the structure and properties (governing parameters) of the bifurcation model of global climate on different time scales.
Since the bifurcation model describes the climatic trends of paleotemperature and global ice volume well without consideration of the mechanism of solar power pacemaker, the natural question arises here: "Is it possible that this fact contradicts the assigned task?" Below we will show that there is no contradiction here, because actually the mechanism of solar power pacemaker is implicitly taken into account in the climatic potential (\ref{eq7}), and it is non-trivial confirmation of significance of the principle of structural invariance.

\section{Axion mechanism of solar dynamo - geodynamo connection}

We have shown that strong correlation between the temporal variations of magnetic field of the Earth (Y-component) and the magnetic field toroidal component of tachocline zone of the Sun really takes place. Thereupon we have asked ourselves: "May hypothetical solar axions\footnote{Axion models are motivated by the strong CP problem - the apparent vanishing of the CP- and T-violating electrical dipole moment (EDM) of the neutron. The axion model offers a dynamical solution to the strong CP problem by introducing a new scalar field which rolls within its potential into a state of minimum action, a CP-conserving QCD vacuum state. Any imbalance between the contributions to the EDM from TeV and GeV scales is absorbed into the scalar field value. The quantized excitations of the scalar field about the potential minimum are called axions (see \cite{ref11} and refs. therein).}, which can transform into photons in external electric or magnetic field (the so-called inverse Primakoff effect), be the instrument by which the magnetic field of the solar tachocline zone modulates the magnetic field of the Earth? In other words, may solar axions be an effective energy source and modulator of the Earth magnetic field?"

\begin{figure*}
  \includegraphics{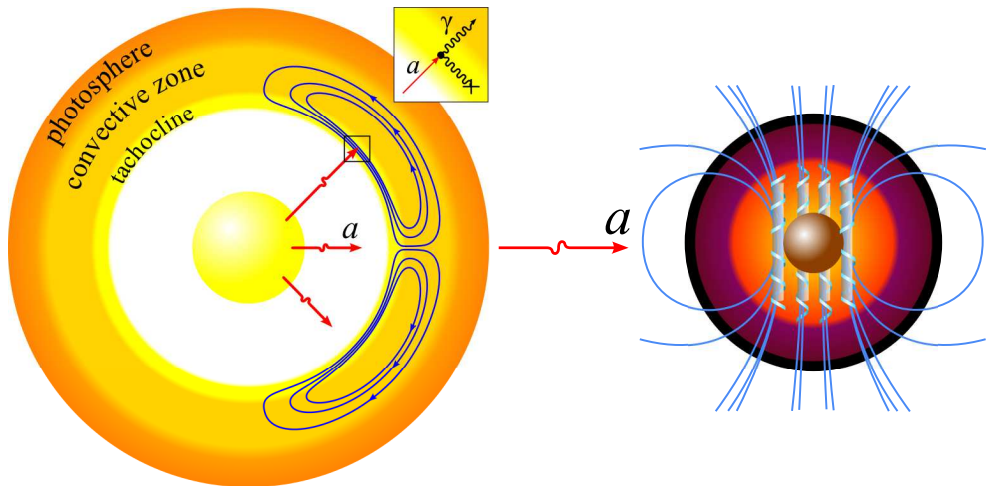}
  \caption{\label{fig4}Schematic picture of the solar tachocline zone, Earth's liquid outer (red region) and inner (brown region) core. Solar axions are resonance absorbed in iron of the Earth core conversing into $\gamma$-quanta, which are the supplementary energy source in the Earth core. Blue lines on the Sun designate the magnetic field. Note: In the conventional concept, the molten iron of liquid phase of Earth's core circulates along a spiraling in columns aligned in the north-south direction, generating electrical currents that set up the dipolar magnetic field. The concentration of field lines into anticyclonic vortices (rotating in the same as air around a region of high pressure) has been thought to explain the intense magnetic lobes found in Earth's field at the top of the core.}
\end{figure*}

It turns out that it is really possible \cite{ref8}. Following \cite{ref8}, let us consider without loss of generality the simplified axion mechanism of solar dynamo-geodynamo connection. As is known, the reaction of the solar cycle that produces solar energy is one of axion sources. Since axions are pseudoscalar particles they can be emitted in nuclear magnetic transitions. On the other hand, since the temperature in the center of the Sun is 1.3 keV, some nuclei having low-lying nuclear level can be excited thermally. At the same time, monochromatic axions can be emitted in the nuclear magnetic transitions from the first thermally excited level to the ground state. Below we consider only 14.4 keV solar axions emitted by the M1 transition in $^{57}$Fe nuclei, because just these axions can be resonantly absorbed in the iron core of the Earth generating 14.4 keV $\gamma$-quanta by the discharge of the excited nuclear level (Figure \ref{fig4}).

It is interesting, that exactly these $\gamma$-quanta with the energy 14.4 keV are the supplementary energy source in the Earth core, which can pretend to the role of energy source of generation and modulator of the Earth magnetic field. At the same time there is a natural question, is this energy sufficient for generation of the magnetic field of the Earth and how this source can execute the role of the modulator of the Earth magnetic field. To answer these questions let us briefly consider the axion "course of life" inside the Sun before it leaves the Sun. It appears \cite{ref8} that passing through the solar tachocline zone (the bottom of the Sun convective zone), where the Sun magnetic field is generated, axions can be converted into $\gamma$-quanta and thereby to decrease the solar axions flux to the Earth. As is shown in \cite{ref8}, in this case the probability that an axion converts back to a "observable" photon inside the magnetic field can be represented by the following simple form

\begin{equation}
  \label{eq8}
  P_{a \gamma} \cong \left( \dfrac{g_{a \gamma BL}}{2} \right)^2,
\end{equation}

\noindent where $g_{a \gamma} \sim 1.64 \cdot 10^9~GeV^{-1}$ is the strength of an axion coupling to a photon, $L \sim 3.5 \cdot 10^{-7}~m$ is the thickness of solar tachocline zone, $B \sim 35~T$ is the conservative value for the magnetic field of the active Sun. From this it follows that the solar axion flux outgoing beyond the Sun is modulated by the value of the Sun magnetic field (see (\ref{eq7})). At the same time, it is obvious, that the axion flux to the Earth is low during the active Sun and conversely it is practically maximal during the quiet Sun.

Now let us show that the total energy of axions during the quiet Sun is sufficient to generate the Earth magnetic field. It is not difficult to show \cite{ref8} that the axion resonant absorption rate in the Earth core, which contains the $N_{Fe}^{57}$ nuclei of $^{57}Fe$ isotope, is about

\begin{equation}
  \label{eq9}
  R_a \approx 5.2 \cdot 10^{-3} \left( g_{aN}^{eff} \right)^4 N_{Fe}^{57} \left[ 1 - P_{a \rightarrow \gamma} \right],
\end{equation}

\noindent where

\begin{equation}
  \label{eq10}
  P_{a \rightarrow \gamma} \sim 
  \begin{cases}
    1 \text{ at } B_{ST} \approx 35T \\
    0 \text{ at } B_{ST} \leq 5.0T  
  \end{cases}.
\end{equation}

It is known, that the number of $^{57}Fe$ nuclei in the Earth core is $N_{Fe}^{57} \sim 3 \cdot 10^{47}$ \cite{ref8} and the average energy of $^{57}Fe$ solar axions is $ \left\langle E_a \right\rangle$= 14.4 keV. If in (\ref{eq9}) for an axion-nucleon coupling $g_{aN}^{eff} \sim 10^{-5}$ \cite{ref8} to take into account the factor 2 related to uncertainty of iron concentration profile at the Sun, then with an allowance of (\ref{eq9}) the maximum energy release rate $W_{\gamma}$ in the Earth core is equal to

\begin{equation}
  \label{eq11}
  W_{\gamma} = R_a \cdot \left\langle E_a \right\rangle \sim 1~TW.
\end{equation}

Analysis of modern model parameters of thermal state of the Earth's core \cite{ref1} shows that in spite of known difficulties in interpretation of the results of evolutionary geodynamo simulation, such a thermal power (1 TW) is sufficient for generation and maintenance of the Earth magnetic field \cite{ref1}. It is easy to show that it is exactly so by the known dependence of magnetic field $B_E$ on the total ohmic dissipation $D$ in the Earth core 

\begin{equation}
  \label{eq12}
  D \sim \dfrac{\eta \cdot V}{\mu \cdot d_B^2} B_E^2,
\end{equation}

\noindent where $\eta$ is the magnetic diffusivity, $V = (4/3)\pi r_{core}^3 $is the core volume, $\mu$ is the permeability, $d_B$ is the characteristic length scale on which the field vector changes. If consider that $\eta \sim 1~m^2 / s$, $r_{core} \sim d_B$ and $\mu \sim 1$, in the case $D \sim W_{\gamma} \sim 1~TW$ we obtain the value of toroidal magnetic field $B_E \sim 0.3~T$, which is in good agreement with theoretical estimations \cite{ref8}.

At the same time, in spite of the fact that the axion mechanism of solar dynamo-geodynamo connection explains well the strong negative correlation between  the magnetic field of the solar tachocline zone and the Earth magnetic field from the physical standpoint, it can not explain other correlations in Figure \ref{fig1} (between the magnetic field of the solar tachocline zone and variations of the Earth angular velocity, average global ocean level and the number of large earthquakes with the magnitude M$\geq$7) from the energy standpoin. However, under certain conditions, i.e. within the framework of the hypothesis of natural nuclear georeactor existence on the boundary of the liquid and solid phases of the Earth core \cite{ref15,ref16}, the axion mechanism can effectively provide these correlations

\section{Soliton-like nuclear georeactor and axion mechanism of the Earth core "heating"}

Now it is obvious that the magnificent experiments of the KamLAND-collobaration over the last 8 years \cite{ref17} have been extremely important not only for observation of reactor antineutrino oscillations. They make it possible for the first time to verify one of most vivid and mysterious ideas in nuclear geophysics – the hypothesis of natural nuclear georeactor existence (see \cite{ref15} and refs. therein). In spite of its singularity and long history, this hypothesis becomes especially attractive today because it enables to explain clearly from the physical standpoint different unrelated, at the first glance, geophysical anomalous phenomena whose fundamental nature is beyond doubt \cite{ref18}.

\begin{figure}
  \includegraphics{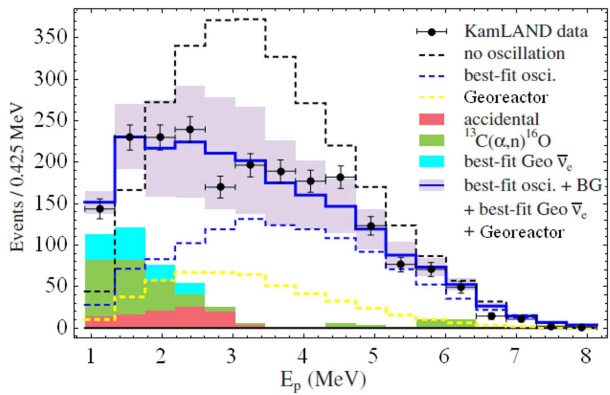}
  \caption{\label{fig5}Prompt event energy spectrum of $\tilde{\nu}_e $ candidate events (the years 2002-2009) \cite{ref16}. The shaded background and geoneutrino histograms are cumulative. Statistical uncertainties are shown for the data; the violet band on the blue histogram indicates the event rate systematic uncertainty within the framework of the georeactor hypothesis. The total georeactor power is 29.7 TW. Georeactors are at a distance of 6400 and 6830 km from the KamLAND-detector.}
\end{figure}

\begin{figure*}
  \includegraphics[width=0.75\linewidth]{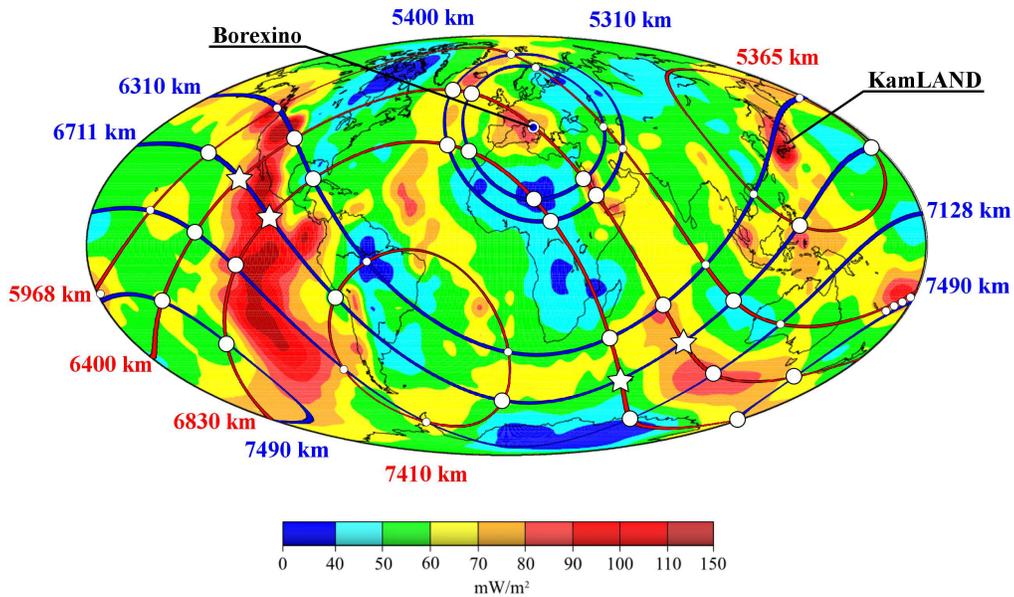}
  \caption{\label{fig6}Distribution of geothermal power density on the Earth \cite{ref21} superposed with the conjugate "pseudogeoreactor" ellipsoidal closed curves, which were built on basis of KamLAND (red lines) and Borexino (blue lines) experimental data \cite{ref16}. ($\star$)– operating nuclear georeactors; ($\bigcirc$) and ($\circ$) – nuclear georeactors, whose power (if they are operating) is an order of magnitude or more less than the thermal power of reactors designated by ($\star$).}
\end{figure*}

We have to note that, in spite of the fact that the experimental KamLAND-data are well described within the framework of georeactor model \cite{ref15,ref16} (see Figure \ref{fig5}) and the location of soliton-like nuclear georeactors (Figure \ref{fig6} \cite{ref16}) is determined by triangulation of the KamLAND \cite{ref17} and Borexino data \cite{ref18}, some geophysicists have doubts not only about existence of the georeactor, but in the first place about its power. In this connection we would like to pay attention to the strange restriction (W $\leq$ 6.2 TW) on the value of nuclear georeactor thermal power W, which, unfortunately, has been frequently met in the scientific literature recently \cite{ref18,ref19,ref20,ref22}. This restriction terrifically masks and distorts clear understanding of the problem of georeactor existence, which is intricate enough by itself.

Indeed, one of the conclusions of the KamLAND–colloboration is the upper bound of nuclear georeactor thermal power (W $\leq$ 6.2 TW at 90\% C.L.), which is a direct consequence of uncertainty of KamLAND experimental data \cite{ref19}. However, it is necessary to keep firmly in mind that this restriction is true only for the concrete parameters of mixing ($\Delta m_{21}^2 = 7.58 \cdot 10^{-5}~eV^2, \tan{\theta_{12}} = 0.56$) obtained within the framework of the concrete $\chi^2$-hypothesis of KamLAND-experiment which takes into account the existence of georeactor within the framework of nonzero hypothesis \cite{ref20}, but absolutely ignores such a nontrivial property of the nuclear georeactor as an uncertainty of georeactor antineutrino spectrum, which in the case of soliton-like nuclear georeactor reaches $\sim$100\% \cite{ref16}. As shown in Ref. \cite{ref16}, the account of this uncertainty within the framework of maximum likelihood function leads (in the minimization of the $\chi^2$-function)  to  considerable expansion of restriction on the nuclear georeactor heat power ($\sim$30 TW) and, accordingly, to the new oscillation parameters ($\Delta m_{21}^2 = 2.5 \cdot 10^{-5}~eV^2, \tan{\theta_{12}} = 0.437$) for reactor antineutrino.

However, in spite of obvious attractiveness of the hypothesis of natural nuclear georeactor existence there are some difficulties for its perception predetermined by non-trivial properties which georeactor must possess. At first, natural, i.e. unenriched, uranium or thorium must be used as a nuclear fuel. Secondly, traditional control rods are completely absent in the reactivity regulation system of reactor. Thirdly, in spite of the absence of control rods a reactor must possess the property of so-called inner safety. It means that the critical state of the reactor core must be permanently maintained in any situation, i.e. normal operation of the reactor is automatically maintained not as a result of operator's activity, but by virtue of physical reasons-laws preventing the explosive development of chain reaction by natural way \cite{ref24}. Figuratively speaking, the reactor with inner safety is the "nuclear installation which never explodes" \cite{ref25}.

It seems to be strange, but reactors satisfying such unusual requirements are possible in reality. For the first time the idea of such a self-regulating fast reactor (so-called mode of breed-and-burn) was expressed in a general form by Russian physicists Feynberg and Kunegin \cite{ref26} and relatively recently "reanimated" as an idea of the self-regulating fast reactor in traveling-wave mode of nuclear burning by L. Feoktistov \cite{ref27} and independently by Teller, Ishikawa and Wood \cite{ref28}. 

\begin{figure*}
  \includegraphics{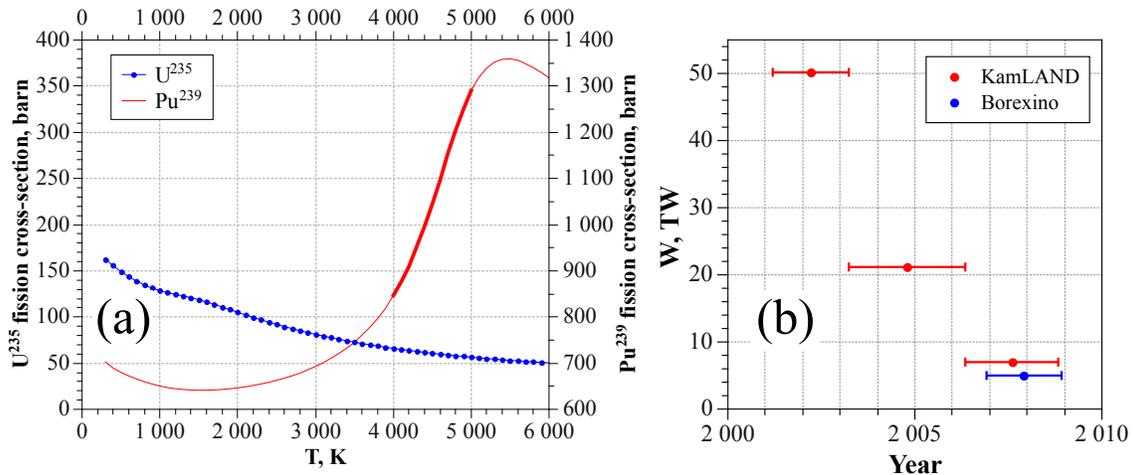}
  \caption{\label{fig7}(a) Dependence of the $^{239}$Pu fission cross-section averaged over the neutron spectrum on fuel medium temperature for limiting energy (3kT) of the Fermi and Maxwell spectra. The similar dependence for the $^{235}$U fission cross-section is shown for comparison. (b) Time evolution of the nuclear georeactor thermal power W.}
\end{figure*}

The discussed nuclear georeactor located on the boundary of the liquid and solid phases of the Earth core has an unique and important for our aims feature, which consists in the fact that the fission cross-section of $^{239}$Pu (generated due to the georeactor operating) is the sharply nonlinear function of temperature in the range 3000-5000 K (Figure \ref{fig7}a). It means that the variations of the Earth core temperature generated by the mechanism of solar dynamo-geodynamo connection will induce corresponding variations of the nuclear georeactor thermal power. It's strange, but it is true and it is confirmed by inverse correlation between the solar magnetic field (Figure \ref{fig1}) and the nuclear georeactor thermal power over the period 2002-2009 (Figure \ref{fig7}b). Thus, on the one hand, such a coordinated behavior of the solar magnetic field and the nuclear georeactor thermal power is an indirect confirmation of reality of the "axion mechanism-nuclear georeactor" energy chain and, on the other hand, according to the estimated power of various geophysical processes \cite{ref29}, such a generalized mechanism can provide the solar-terrestrial correlations shown in Figure \ref{fig1} effectively.

If the georeactor hypothesis is true, the fluctuations of georeactor thermal power can influence the Earth's global climate in the form of anomalous temperature jumps in the following way. Strong fluctuations of the georeactor thermal power can lead to partial blocking of convection in the liquid core \cite{ref10} and change of angular velocity of liquid geosphere rotation, thereby, by virtue of the conservation law of Earth's angular moment to change of angular velocity of the mantle and the Earth's surface respectively. It means that the heat or, more precisely, dissipation energy caused by friction of earthly surface and bottom layer can make a considerable contribution to the total energy balance of the atmosphere and thereby significantly influence on the Earth global climate evolution \cite{ref9,ref10}.

\section{Bifurcation model of the Earth global climate on the annual time scale}

Newton's second law for friction rough surfaces (the Earth surface and the atmosphere surface layer) with an allowance for nonlinear friction by Gilmore \cite{ref30} and the climatic potential (\ref{eq7}) has the form of the van der Pole-Duffing type equation (see Figure \ref{fig2}):

\begin{equation}
  \label{eq13}
  m \ddot{x} = - \mu \left( x^2 - \lambda \right) \dot{x} - \partial_x U,
\end{equation}

\noindent where $x$ is the average shift length of the atmosphere surface layer relative to the Earth surface, $m$ is the effective mass of boundary layer, $\mu$ and $\lambda$ are parameters, $U$ is the climatic potential of (\ref{eq7}) type.

Using the substitutions $x = \Delta \omega \cdot R \cdot \Delta t$, $\xi = \omega R$, $\nu = \Delta \omega /\omega$, we can write down (\ref{eq13}) in the following form 

\begin{equation}
  \label{eq14}
  \xi m \dot{\nu} = - \mu \left[ \xi^2 \left( \Delta t \right)^2 \nu^2 - \lambda \right] \xi \nu - \dfrac{1}{\xi \Delta t} \dfrac{\partial U}{\partial \nu},
\end{equation}

\noindent where $\omega$ is the angular velocity of the Earth rotation and $\Delta \omega$ is its change over $\Delta t$= 1 year, $R$ is the average Earth radius, $\nu$ is the dimensionless quantity which describes by definition the Earth rotational velocity \cite{ref5} (Figure \ref{fig1}).

Since temporal variations of the global ocean level and temporal variations of the Earth average temperature strongly correlate without time lag, whereas the temporal variations of the global ocean level and temporal variations of the Earth rotational velocity strongly correlate with the lag $t_{lag} \sim$5 years (see Figure \ref{fig1}), (\ref{eq14}) with consideration of the approximate equality $\nu \sim kT_{t-t_{lag}}$ can be rewritten in the following form

\begin{equation}
  \label{eq15}
  \xi k m \dot{T} = - \mu \xi k \left[ \xi^2 k^2 \left( \Delta t \right)^2 T^2  - \lambda \right] T - \dfrac{1}{\xi k \Delta t} \dfrac{\partial U}{\partial T},
\end{equation}

\noindent that explicitly takes into account the mechanism of solar power pacemaker.

\begin{figure}
  \includegraphics{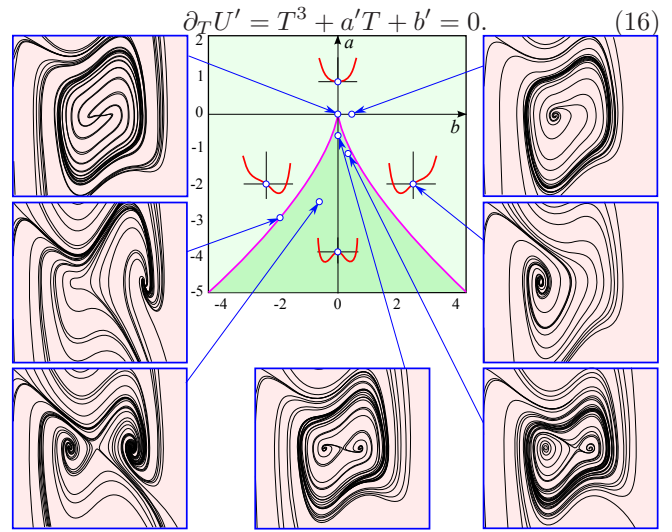}
  \caption{\label{fig8}A plane of the parameters (a, b), the typical shapes of the assembly-type catastrophe potential (red lines) and the phase portraits (black lines on the pink squares) of the self-oscillating system of the van der Pole-Duffing type (\ref{eq13}) at $m, \lambda =1$. Blue circles are points, to whose coordinates the phase portraits and the assembly-type catastrophe potential correspond.}
\end{figure}

Nontrivial properties of the basic equation of bifurcation model of the Earth global climate on the annual time scale are exhibited in Figure \ref{fig8} by the variety of phase portraits depending on the governing parameters ($a, b$). Moreover, the change of shape of the assembly-type catastrophe potentials (\ref{eq7}) on the plane ($a, b$) directly specifies the conditions of "warm-cold" phase transitions in the climatic self-oscillatory system of the van der Pole-Duffing type (\ref{eq13}).

Here it is interesting to note the following remarkable fact. It was found that the low order dynamic models of the time evolution of the toroidal magnetic field of the Sun derived from mean field dynamo theory are also described by the nonlinear oscillator equations of the van der Pole-Duffing type \cite{ref31,ref32}. In this sense, the identical type of equations describing the time evolution both of the Sun magnetic field and the Earth global climate is one more confirmation and, at the same time, natural consequence of physical, i.e., really existing, mechanism of solar dynamo-geodynamo connection.

Now we return to the problem of taking into account of the mechanism of solar power pacemaker within the framework of the bifurcation model of the Earth global climate on different time scales. It is known that on the large time scales (from several to ten thousands years) on which our bifurcation model was considered above, the equilibrium state of the global climate is reached at every time point. It is obvious, that in this case the left-hand side of (\ref{eq15}) can be set equal to zero, and (\ref{eq7}) itself can be written down in the following form

\begin{equation}
  \label{eq16}
  \partial_T U' = T^3 + a'T + b' = 0.
\end{equation}

It means that the bifurcation model of the Earth global climate on the ten thousandth time scale really takes into account not only the laws of atmospheric physics, in particular, the laws of geometrical optics of climatic billiards, which generalize the cosmic rays-clouds effect by Sven-smark, the first (the Twomey effect) and second indirect aerosol effects \cite{ref9}, but also the mechanism of solar power pacemaker, which was masked before \cite{ref9,ref10} by renormalization procedure of the governing parameters to take into account the initial conditions. In other words, theoretical solutions of the bifurcation model of the Earth global climate on the ten thousandth time scale with respect to the temperature and the global ice volume not only take into account the mechanism of solar dynamo-geodynamo connection but, in combination with high quality of description of the known experimental trends of the temperature and the global ice volume \cite{ref9,ref10}, are reliable confirmation of correct and holistic understanding of the basic foundations of nonlinear physics of the Earth global climate formation.

\bibliography{PacemakerLit}

\begin{thebibliography}{33}%
\makeatletter
\providecommand \@ifxundefined [1]{%
 \@ifx{#1\undefined}
}%
\providecommand \@ifnum [1]{%
 \ifnum #1\expandafter \@firstoftwo
 \else \expandafter \@secondoftwo
 \fi
}%
\providecommand \@ifx [1]{%
 \ifx #1\expandafter \@firstoftwo
 \else \expandafter \@secondoftwo
 \fi
}%
\providecommand \natexlab [1]{#1}%
\providecommand \enquote  [1]{``#1''}%
\providecommand \bibnamefont  [1]{#1}%
\providecommand \bibfnamefont [1]{#1}%
\providecommand \citenamefont [1]{#1}%
\providecommand \href@noop [0]{\@secondoftwo}%
\providecommand \href [0]{\begingroup \@sanitize@url \@href}%
\providecommand \@href[1]{\@@startlink{#1}\@@href}%
\providecommand \@@href[1]{\endgroup#1\@@endlink}%
\providecommand \@sanitize@url [0]{\catcode `\\12\catcode `\$12\catcode
  `\&12\catcode `\#12\catcode `\^12\catcode `\_12\catcode `\%12\relax}%
\providecommand \@@startlink[1]{}%
\providecommand \@@endlink[0]{}%
\providecommand \url  [0]{\begingroup\@sanitize@url \@url }%
\providecommand \@url [1]{\endgroup\@href {#1}{\urlprefix }}%
\providecommand \urlprefix  [0]{URL }%
\providecommand \Eprint [0]{\href }%
\providecommand \doibase [0]{http://dx.doi.org/}%
\providecommand \selectlanguage [0]{\@gobble}%
\providecommand \bibinfo  [0]{\@secondoftwo}%
\providecommand \bibfield  [0]{\@secondoftwo}%
\providecommand \translation [1]{[#1]}%
\providecommand \BibitemOpen [0]{}%
\providecommand \bibitemStop [0]{}%
\providecommand \bibitemNoStop [0]{.\EOS\space}%
\providecommand \EOS [0]{\spacefactor3000\relax}%
\providecommand \BibitemShut  [1]{\csname bibitem#1\endcsname}%
\let\auto@bib@innerbib\@empty
\bibitem [{\citenamefont {Buffet}(2003)}]{ref1}%
  \BibitemOpen
  \bibfield  {author} {\bibinfo {author} {\bibfnamefont {B.~A.}\ \bibnamefont
  {Buffet}},\ }\href@noop {} {\bibfield  {journal} {\bibinfo  {journal}
  {Science}\ }\textbf {\bibinfo {volume} {299}},\ \bibinfo {pages} {1675}
  (\bibinfo {year} {2003})}\BibitemShut {NoStop}%
\bibitem [{Note1()}]{Note1}%
  \BibitemOpen
  \bibinfo {note} {Note that the strong (negative) correlation between the
  temporal variations of magnetic flux in the tachocline zone and the Earth
  magnetic field (Y-component) will be observed only for experimental data
  obtained at that observatories where the temporal variations of declination
  ($\delta D / \delta t$) or the closely associated east component ($\delta Y /
  \delta t$) are directly proportional to the westward drift of magnetic
  features \cite {ref2}. This condition is very important for understanding of
  physical nature of indicated above correlation, so far as it is known that
  just motions of the top layers of the Earth's core are responsible for most
  magnetic variations and, in particular, for the westward drift of magnetic
  features seen on the Earth's surface on the decade time scale. Europe and
  Australia are geographical places, where this condition is fulfilled (see
  Figure 2 in \cite {ref2}).}\BibitemShut {Stop}%
\bibitem [{\citenamefont {Mouel}\ \emph {et~al.}(1981)\citenamefont {Mouel},
  \citenamefont {Madden}, \citenamefont {Ducruix},\ and\ \citenamefont
  {Courtillot}}]{ref2}%
  \BibitemOpen
  \bibfield  {author} {\bibinfo {author} {\bibfnamefont {J.-L.~L.}\
  \bibnamefont {Mouel}}, \bibinfo {author} {\bibfnamefont {T.~R.}\ \bibnamefont
  {Madden}}, \bibinfo {author} {\bibfnamefont {J.}~\bibnamefont {Ducruix}}, \
  and\ \bibinfo {author} {\bibfnamefont {V.}~\bibnamefont {Courtillot}},\
  }\href@noop {} {\bibfield  {journal} {\bibinfo  {journal} {Nature}\ }\textbf
  {\bibinfo {volume} {290}},\ \bibinfo {pages} {763} (\bibinfo {year}
  {1981})}\BibitemShut {NoStop}%
\bibitem [{\citenamefont {Dikpati}\ \emph {et~al.}(2008)\citenamefont
  {Dikpati}, \citenamefont {de~Toma},\ and\ \citenamefont {Gilman}}]{ref3}%
  \BibitemOpen
  \bibfield  {author} {\bibinfo {author} {\bibfnamefont {M.}~\bibnamefont
  {Dikpati}}, \bibinfo {author} {\bibfnamefont {G.}~\bibnamefont {de~Toma}}, \
  and\ \bibinfo {author} {\bibfnamefont {P.~A.}\ \bibnamefont {Gilman}},\
  }\href@noop {} {\bibfield  {journal} {\bibinfo  {journal} {The Astrophysics
  Journal}\ }\textbf {\bibinfo {volume} {675}},\ \bibinfo {pages} {920}
  (\bibinfo {year} {2008})}\BibitemShut {NoStop}%
\bibitem [{ref(2007)}]{ref4}%
  \BibitemOpen
  \href@noop {} {\emph {\bibinfo {title} {Data of the observatory Eskdalemuir
  (England)}}},\ \bibinfo {type} {Tech. Rep.}\ (\bibinfo  {institution} {World
  Data Centre for Geomagnetic (Edinburg)},\ \bibinfo {year} {2007})\ \bibinfo
  {note} {worldwide
  \url{http://www.geomag.bgs.ak.uk./gifs/annual_means.shtml}}\BibitemShut
  {NoStop}%
\bibitem [{\citenamefont {Rusov}\ \emph {et~al.}({\natexlab{a}})\citenamefont
  {Rusov}, \citenamefont {Linnik}, \citenamefont {Kudela}, \citenamefont
  {Mavrodiev}, \citenamefont {Sharph}, \citenamefont {Zelentsova},
  \citenamefont {Beglaryan}, \citenamefont {Smolyar},\ and\ \citenamefont
  {Merkotan}}]{ref8}%
  \BibitemOpen
  \bibfield  {author} {\bibinfo {author} {\bibfnamefont {V.~D.}\ \bibnamefont
  {Rusov}}, \bibinfo {author} {\bibfnamefont {E.~P.}\ \bibnamefont {Linnik}},
  \bibinfo {author} {\bibfnamefont {K.}~\bibnamefont {Kudela}}, \bibinfo
  {author} {\bibfnamefont {S.~C.}\ \bibnamefont {Mavrodiev}}, \bibinfo {author}
  {\bibfnamefont {I.~V.}\ \bibnamefont {Sharph}}, \bibinfo {author}
  {\bibfnamefont {T.~N.}\ \bibnamefont {Zelentsova}}, \bibinfo {author}
  {\bibfnamefont {M.~E.}\ \bibnamefont {Beglaryan}}, \bibinfo {author}
  {\bibfnamefont {V.~P.}\ \bibnamefont {Smolyar}}, \ and\ \bibinfo {author}
  {\bibfnamefont {K.~K.}\ \bibnamefont {Merkotan}},\ }\href@noop {} {\enquote
  {\bibinfo {title} {Axion mechanism of the sun luminosity and solar dynamo -
  geodynamo connection},}\ } ({\natexlab{a}}),\ \bibinfo {note}
  {arXiv:1009.3340}\BibitemShut {NoStop}%
\bibitem [{\citenamefont {Sidorenkov}(2009)}]{ref5}%
  \BibitemOpen
  \bibfield  {author} {\bibinfo {author} {\bibfnamefont {N.~S.}\ \bibnamefont
  {Sidorenkov}},\ }\href@noop {} {\emph {\bibinfo {title} {The Interaction
  Between Earth's Rotation and Geophysical Processes}}}\ (\bibinfo
  {publisher} {Wiley-VCH},\ \bibinfo {year} {2009})\BibitemShut {NoStop}%
\bibitem [{ref()}]{ref6}%
  \BibitemOpen
  \href@noop {} {\enquote {\bibinfo {title} {Pacific decade-oscillation
  (pdo)+atlantic multidecaded oscillation (oma)},}\ }\bibinfo {howpublished}
  {Internet Available
  \url{http:/www.appinsys.com/GlobalWarming/PDO_AMO.htm}}\BibitemShut {NoStop}%
\bibitem [{\citenamefont {Engdahi}\ and\ \citenamefont
  {Villsenor}(2002)}]{ref7}%
  \BibitemOpen
  \bibfield  {author} {\bibinfo {author} {\bibfnamefont {E.~R.}\ \bibnamefont
  {Engdahi}}\ and\ \bibinfo {author} {\bibfnamefont {A.}~\bibnamefont
  {Villsenor}},\ }\enquote {\bibinfo {title} {Global seismicity; 1990-1999},}\
  \ (\bibinfo  {publisher} {Academic Press},\ \bibinfo {year} {2002})\ Chap.\
  \bibinfo {chapter} {Part A (International Geophysics)}\BibitemShut {NoStop}%
\bibitem [{\citenamefont {Rusov}\ \emph
  {et~al.}(2010{\natexlab{a}})\citenamefont {Rusov}, \citenamefont {Glushkov},
  \citenamefont {Vaschenko}, \citenamefont {Myhalus}, \citenamefont
  {Bondartchuk}, \citenamefont {Smolyar}, \citenamefont {Linnik}, \citenamefont
  {Mavrodiev},\ and\ \citenamefont {Vachev}}]{ref9}%
  \BibitemOpen
  \bibfield  {author} {\bibinfo {author} {\bibfnamefont {V.~D.}\ \bibnamefont
  {Rusov}}, \bibinfo {author} {\bibfnamefont {A.~V.}\ \bibnamefont {Glushkov}},
  \bibinfo {author} {\bibfnamefont {V.~N.}\ \bibnamefont {Vaschenko}}, \bibinfo
  {author} {\bibfnamefont {O.~T.}\ \bibnamefont {Myhalus}}, \bibinfo {author}
  {\bibfnamefont {Y.~A.}\ \bibnamefont {Bondartchuk}}, \bibinfo {author}
  {\bibfnamefont {V.~P.}\ \bibnamefont {Smolyar}}, \bibinfo {author}
  {\bibfnamefont {E.~P.}\ \bibnamefont {Linnik}}, \bibinfo {author}
  {\bibfnamefont {S.~C.}\ \bibnamefont {Mavrodiev}}, \ and\ \bibinfo {author}
  {\bibfnamefont {B.~I.}\ \bibnamefont {Vachev}},\ }\href@noop {} {\bibfield
  {journal} {\bibinfo  {journal} {J. Atmos. Sol.-Terr. Phys.}\ }\textbf
  {\bibinfo {volume} {72}},\ \bibinfo {pages} {398} (\bibinfo {year}
  {2010}{\natexlab{a}})},\ \bibinfo {note} {arXiv: physics.ao-ph
  0803.2765}\BibitemShut {NoStop}%
\bibitem [{\citenamefont {Rusov}\ \emph
  {et~al.}(2010{\natexlab{b}})\citenamefont {Rusov}, \citenamefont {Vaschenko},
  \citenamefont {Linnik}, \citenamefont {Myhalus}, \citenamefont {Bondartchuk},
  \citenamefont {Smolyar}, \citenamefont {Kosenko}, \citenamefont {Mavrodiev},\
  and\ \citenamefont {Vachev}}]{ref10}%
  \BibitemOpen
  \bibfield  {author} {\bibinfo {author} {\bibfnamefont {V.~D.}\ \bibnamefont
  {Rusov}}, \bibinfo {author} {\bibfnamefont {V.~N.}\ \bibnamefont
  {Vaschenko}}, \bibinfo {author} {\bibfnamefont {E.~P.}\ \bibnamefont
  {Linnik}}, \bibinfo {author} {\bibfnamefont {Ð.~T.}\ \bibnamefont {Myhalus}},
  \bibinfo {author} {\bibfnamefont {Y.~A.}\ \bibnamefont {Bondartchuk}},
  \bibinfo {author} {\bibfnamefont {V.~P.}\ \bibnamefont {Smolyar}}, \bibinfo
  {author} {\bibfnamefont {S.~.}\ \bibnamefont {Kosenko}}, \bibinfo {author}
  {\bibfnamefont {S.~C.}\ \bibnamefont {Mavrodiev}}, \ and\ \bibinfo {author}
  {\bibfnamefont {B.~I.}\ \bibnamefont {Vachev}},\ }\href@noop {} {\bibfield
  {journal} {\bibinfo  {journal} {J. Atmos. Sol.-Terr. Phys.}\ }\textbf
  {\bibinfo {volume} {72}},\ \bibinfo {pages} {389} (\bibinfo {year}
  {2010}{\natexlab{b}})},\ \bibinfo {note} {arXiv: physics.ao-ph
  0803.2766}\BibitemShut {NoStop}%
\bibitem [{\citenamefont {Bassinot}\ \emph {et~al.}(1994)\citenamefont
  {Bassinot}, \citenamefont {Labeyrie}, \citenamefont {Vincent}, \citenamefont
  {Quidelleur}, \citenamefont {Shackleton},\ and\ \citenamefont
  {Lancelot}}]{ref12}%
  \BibitemOpen
  \bibfield  {author} {\bibinfo {author} {\bibfnamefont {F.~C.}\ \bibnamefont
  {Bassinot}}, \bibinfo {author} {\bibfnamefont {L.~D.}\ \bibnamefont
  {Labeyrie}}, \bibinfo {author} {\bibfnamefont {E.}~\bibnamefont {Vincent}},
  \bibinfo {author} {\bibfnamefont {X.}~\bibnamefont {Quidelleur}}, \bibinfo
  {author} {\bibfnamefont {N.~J.}\ \bibnamefont {Shackleton}}, \ and\ \bibinfo
  {author} {\bibfnamefont {Y.}~\bibnamefont {Lancelot}},\ }\href@noop {}
  {\bibfield  {journal} {\bibinfo  {journal} {Earth Planet. Sci. Lett.}\
  }\textbf {\bibinfo {volume} {126}},\ \bibinfo {pages} {91} (\bibinfo {year}
  {1994})}\BibitemShut {NoStop}%
\bibitem [{\citenamefont {Imbrie}\ \emph {et~al.}(1993)\citenamefont {Imbrie},
  \citenamefont {Berger}, \citenamefont {Boyle}, \citenamefont {Clemens},
  \citenamefont {Duffy}, \citenamefont {Howard}, \citenamefont {Kukla},
  \citenamefont {Kutzbach}, \citenamefont {Martinson}, \citenamefont
  {McIntyre}, \citenamefont {Mix}, \citenamefont {Molfino}, \citenamefont
  {Morley}, \citenamefont {Peterson}, \citenamefont {adn W.~L.~Prell},
  \citenamefont {Raymo}, \citenamefont {Shackleton},\ and\ \citenamefont
  {Toggweiler}}]{ref13}%
  \BibitemOpen
  \bibfield  {author} {\bibinfo {author} {\bibfnamefont {J.}~\bibnamefont
  {Imbrie}}, \bibinfo {author} {\bibfnamefont {A.}~\bibnamefont {Berger}},
  \bibinfo {author} {\bibfnamefont {E.~A.}\ \bibnamefont {Boyle}}, \bibinfo
  {author} {\bibfnamefont {S.~C.}\ \bibnamefont {Clemens}}, \bibinfo {author}
  {\bibfnamefont {A.}~\bibnamefont {Duffy}}, \bibinfo {author} {\bibfnamefont
  {W.~R.}\ \bibnamefont {Howard}}, \bibinfo {author} {\bibfnamefont
  {G.}~\bibnamefont {Kukla}}, \bibinfo {author} {\bibfnamefont
  {J.}~\bibnamefont {Kutzbach}}, \bibinfo {author} {\bibfnamefont {D.~G.}\
  \bibnamefont {Martinson}}, \bibinfo {author} {\bibfnamefont {A.}~\bibnamefont
  {McIntyre}}, \bibinfo {author} {\bibfnamefont {A.~C.}\ \bibnamefont {Mix}},
  \bibinfo {author} {\bibfnamefont {B.}~\bibnamefont {Molfino}}, \bibinfo
  {author} {\bibfnamefont {J.~J.}\ \bibnamefont {Morley}}, \bibinfo {author}
  {\bibfnamefont {L.~C.}\ \bibnamefont {Peterson}}, \bibinfo {author}
  {\bibfnamefont {N.~G.~P.}\ \bibnamefont {adn W.~L.~Prell}}, \bibinfo {author}
  {\bibfnamefont {M.~E.}\ \bibnamefont {Raymo}}, \bibinfo {author}
  {\bibfnamefont {N.~J.}\ \bibnamefont {Shackleton}}, \ and\ \bibinfo {author}
  {\bibfnamefont {J.~R.}\ \bibnamefont {Toggweiler}},\ }\href@noop {}
  {\bibfield  {journal} {\bibinfo  {journal} {Paleoceanography}\ }\textbf
  {\bibinfo {volume} {8}},\ \bibinfo {pages} {699} (\bibinfo {year} {1993})},\
  \bibinfo {note} {doi:10.1029/93PA02751}\BibitemShut {NoStop}%
\bibitem [{\citenamefont {Tidemann}\ \emph {et~al.}(1994)\citenamefont
  {Tidemann}, \citenamefont {Sarnthein},\ and\ \citenamefont
  {Shackleton}}]{ref14}%
  \BibitemOpen
  \bibfield  {author} {\bibinfo {author} {\bibfnamefont {R.}~\bibnamefont
  {Tidemann}}, \bibinfo {author} {\bibfnamefont {M.}~\bibnamefont {Sarnthein}},
  \ and\ \bibinfo {author} {\bibfnamefont {N.}~\bibnamefont {Shackleton}},\
  }\href@noop {} {\bibfield  {journal} {\bibinfo  {journal} {Paleoceanography}\
  }\textbf {\bibinfo {volume} {9}},\ \bibinfo {pages} {619} (\bibinfo {year}
  {1994})},\ \bibinfo {note} {doi:10.1029/94PA00208}\BibitemShut {NoStop}%
\bibitem [{Note2()}]{Note2}%
  \BibitemOpen
  \bibinfo {note} {Axion models are motivated by the strong CP problem - the
  apparent vanishing of the CP- and T-violating electrical dipole moment (EDM)
  of the neutron. The axion model offers a dynamical solution to the strong CP
  problem by introducing a new scalar field which rolls within its potential
  into a state of minimum action, a CP-conserving QCD vacuum state. Any
  imbalance between the contributions to the EDM from TeV and GeV scales is
  absorbed into the scalar field value. The quantized excitations of the scalar
  field about the potential minimum are called axions (see \cite {ref11} and
  refs. therein).}\BibitemShut {Stop}%
\bibitem [{\citenamefont {Rusov}\ \emph {et~al.}(2007)\citenamefont {Rusov},
  \citenamefont {Pavlovich}, \citenamefont {Vaschenko}, \citenamefont
  {Tarasov}, \citenamefont {Zelentsova}, \citenamefont {Bolshakov},
  \citenamefont {Litvinov}, \citenamefont {Kosenko},\ and\ \citenamefont
  {Byegunova}}]{ref15}%
  \BibitemOpen
  \bibfield  {author} {\bibinfo {author} {\bibfnamefont {V.~D.}\ \bibnamefont
  {Rusov}}, \bibinfo {author} {\bibfnamefont {V.~N.}\ \bibnamefont
  {Pavlovich}}, \bibinfo {author} {\bibfnamefont {V.~N.}\ \bibnamefont
  {Vaschenko}}, \bibinfo {author} {\bibfnamefont {V.~A.}\ \bibnamefont
  {Tarasov}}, \bibinfo {author} {\bibfnamefont {T.~N.}\ \bibnamefont
  {Zelentsova}}, \bibinfo {author} {\bibfnamefont {V.~N.}\ \bibnamefont
  {Bolshakov}}, \bibinfo {author} {\bibfnamefont {D.~A.}\ \bibnamefont
  {Litvinov}}, \bibinfo {author} {\bibfnamefont {S.~I.}\ \bibnamefont
  {Kosenko}}, \ and\ \bibinfo {author} {\bibfnamefont {O.~A.}\ \bibnamefont
  {Byegunova}},\ }\href@noop {} {\bibfield  {journal} {\bibinfo  {journal} {J.
  Geophys. Res.}\ }\textbf {\bibinfo {volume} {112}},\ \bibinfo {pages}
  {B09203} (\bibinfo {year} {2007})},\ \bibinfo {note}
  {doi:10.1029/2005JB004212}\BibitemShut {NoStop}%
\bibitem [{\citenamefont {Rusov}\ \emph {et~al.}({\natexlab{b}})\citenamefont
  {Rusov}, \citenamefont {Litvinov}, \citenamefont {Mavrodiev}, \citenamefont
  {Linnik}, \citenamefont {Vaschenko}, \citenamefont {Zelentsova},
  \citenamefont {Beglaryan}, \citenamefont {Tarasov}, \citenamefont
  {Chernezhenko}, \citenamefont {Smolyar}, \citenamefont {Molchinikolov},\ and\
  \citenamefont {Merkotan}}]{ref16}%
  \BibitemOpen
  \bibfield  {author} {\bibinfo {author} {\bibfnamefont {V.~D.}\ \bibnamefont
  {Rusov}}, \bibinfo {author} {\bibfnamefont {D.~A.}\ \bibnamefont {Litvinov}},
  \bibinfo {author} {\bibfnamefont {S.~C.}\ \bibnamefont {Mavrodiev}}, \bibinfo
  {author} {\bibfnamefont {E.~P.}\ \bibnamefont {Linnik}}, \bibinfo {author}
  {\bibfnamefont {V.~N.}\ \bibnamefont {Vaschenko}}, \bibinfo {author}
  {\bibfnamefont {T.~N.}\ \bibnamefont {Zelentsova}}, \bibinfo {author}
  {\bibfnamefont {M.~E.}\ \bibnamefont {Beglaryan}}, \bibinfo {author}
  {\bibfnamefont {V.~A.}\ \bibnamefont {Tarasov}}, \bibinfo {author}
  {\bibfnamefont {S.~A.}\ \bibnamefont {Chernezhenko}}, \bibinfo {author}
  {\bibfnamefont {V.~P.}\ \bibnamefont {Smolyar}}, \bibinfo {author}
  {\bibfnamefont {P.~O.}\ \bibnamefont {Molchinikolov}}, \ and\ \bibinfo
  {author} {\bibfnamefont {K.~K.}\ \bibnamefont {Merkotan}},\ }\href@noop {}
  {\enquote {\bibinfo {title} {The kamland-experiment and soliton-like
  georeactor. part 1. comparison of theory with experiment},}\ }
  ({\natexlab{b}}),\ \bibinfo {note} {arXiv:1011.3568}\BibitemShut {NoStop}%
\bibitem [{\citenamefont {Gando}\ and\ \citenamefont {others
  (KamLAND~Collaboration)}(2011)}]{ref17}%
  \BibitemOpen
  \bibfield  {author} {\bibinfo {author} {\bibfnamefont {A.}~\bibnamefont
  {Gando}}\ and\ \bibinfo {author} {\bibnamefont {others
  (KamLAND~Collaboration)}},\ }\href@noop {} {\bibfield  {journal} {\bibinfo
  {journal} {Phys. Rev. D}\ }\textbf {\bibinfo {volume} {83}} (\bibinfo {year}
  {2011})},\ \bibinfo {note} {arXiv:1009.4771}\BibitemShut {NoStop}%
\bibitem [{\citenamefont {Bellini}\ and\ \citenamefont {others
  (Borexino~Collaboration)}(2010)}]{ref18}%
  \BibitemOpen
  \bibfield  {author} {\bibinfo {author} {\bibfnamefont {G.}~\bibnamefont
  {Bellini}}\ and\ \bibinfo {author} {\bibnamefont {others
  (Borexino~Collaboration)}},\ }\href@noop {} {\bibfield  {journal} {\bibinfo
  {journal} {Phys. Lett.}\ }\textbf {\bibinfo {volume} {B687}},\ \bibinfo
  {pages} {299} (\bibinfo {year} {2010})}\BibitemShut {NoStop}%
\bibitem [{\citenamefont {Araki}\ and\ \citenamefont {others
  (KamLAND~Collaboration)}(2005)}]{ref19}%
  \BibitemOpen
  \bibfield  {author} {\bibinfo {author} {\bibfnamefont {T.}~\bibnamefont
  {Araki}}\ and\ \bibinfo {author} {\bibnamefont {others
  (KamLAND~Collaboration)}},\ }\href@noop {} {\bibfield  {journal} {\bibinfo
  {journal} {Nature}\ }\textbf {\bibinfo {volume} {436}},\ \bibinfo {pages}
  {499} (\bibinfo {year} {2005})}\BibitemShut {NoStop}%
\bibitem [{\citenamefont {Abe}\ and\ \citenamefont {others
  (KamLAND~Collaboration)}(2008)}]{ref20}%
  \BibitemOpen
  \bibfield  {author} {\bibinfo {author} {\bibfnamefont {S.}~\bibnamefont
  {Abe}}\ and\ \bibinfo {author} {\bibnamefont {others
  (KamLAND~Collaboration)}},\ }\href@noop {} {\bibfield  {journal} {\bibinfo
  {journal} {Phys. Rev. Lett.}\ }\textbf {\bibinfo {volume} {100}},\ \bibinfo
  {pages} {2218031} (\bibinfo {year} {2008})}\BibitemShut {NoStop}%
\bibitem [{\citenamefont {Dye}(2009)}]{ref22}%
  \BibitemOpen
  \bibfield  {author} {\bibinfo {author} {\bibfnamefont {S.~T.}\ \bibnamefont
  {Dye}},\ }\href@noop {} {\bibfield  {journal} {\bibinfo  {journal} {Phys.
  Lett.}\ }\textbf {\bibinfo {volume} {B679}},\ \bibinfo {pages} {15} (\bibinfo
  {year} {2009})}\BibitemShut {NoStop}%
\bibitem [{\citenamefont {Hamza}\ \emph {et~al.}(2008)\citenamefont {Hamza},
  \citenamefont {Cardoso},\ and\ \citenamefont {Neto}}]{ref21}%
  \BibitemOpen
  \bibfield  {author} {\bibinfo {author} {\bibfnamefont {V.~M.}\ \bibnamefont
  {Hamza}}, \bibinfo {author} {\bibfnamefont {R.~R.}\ \bibnamefont {Cardoso}},
  \ and\ \bibinfo {author} {\bibfnamefont {C.~F.~P.}\ \bibnamefont {Neto}},\
  }\href@noop {} {\bibfield  {journal} {\bibinfo  {journal} {International
  Journal of Earth Sciences}\ }\textbf {\bibinfo {volume} {97}},\ \bibinfo
  {pages} {205} (\bibinfo {year} {2008})}\BibitemShut {NoStop}%
\bibitem [{\citenamefont {Rusov}\ \emph {et~al.}(2008)\citenamefont {Rusov},
  \citenamefont {Tarasov},\ and\ \citenamefont {Litvinov}}]{ref24}%
  \BibitemOpen
  \bibfield  {author} {\bibinfo {author} {\bibfnamefont {V.~D.}\ \bibnamefont
  {Rusov}}, \bibinfo {author} {\bibfnamefont {V.~A.}\ \bibnamefont {Tarasov}},
  \ and\ \bibinfo {author} {\bibfnamefont {D.~A.}\ \bibnamefont {Litvinov}},\
  }\href@noop {} {\emph {\bibinfo {title} {Reactor Antineutrinos Physics}}}\
  (\bibinfo  {publisher} {URSS, Moscow},\ \bibinfo {year} {2008})\BibitemShut
  {NoStop}%
\bibitem [{\citenamefont {Feoktistov}(1998)}]{ref25}%
  \BibitemOpen
  \bibfield  {author} {\bibinfo {author} {\bibfnamefont {L.~P.}\ \bibnamefont
  {Feoktistov}},\ }\href@noop {} {\emph {\bibinfo {title} {From the Past
  towards the Future: from the Hopes of Bomb to the Safe Reactor}}}\ (\bibinfo
  {publisher} {Publ. of RFNC-ANRISPh, Snezhinsk, Russia},\ \bibinfo {year}
  {1998})\BibitemShut {NoStop}%
\bibitem [{\citenamefont {Feinberg}(1958)}]{ref26}%
  \BibitemOpen
  \bibfield  {author} {\bibinfo {author} {\bibfnamefont {S.~M.}\ \bibnamefont
  {Feinberg}},\ }in\ \href@noop {} {\emph {\bibinfo {booktitle} {Record of
  Proceedings Session B-10, International Conference on the Peaceful Uses for
  Atomic Energy}}}\ (\bibinfo  {publisher} {United Nations, Geneva,
  Switzerland},\ \bibinfo {year} {1958})\ pp.\ \bibinfo {pages}
  {447--44}\BibitemShut {NoStop}%
\bibitem [{\citenamefont {Feoktistov}(1989)}]{ref27}%
  \BibitemOpen
  \bibfield  {author} {\bibinfo {author} {\bibfnamefont {L.~P.}\ \bibnamefont
  {Feoktistov}},\ }\href@noop {} {\bibfield  {journal} {\bibinfo  {journal}
  {Reports of Academy of Sciences of USSR}\ }\textbf {\bibinfo {volume}
  {309}},\ \bibinfo {pages} {864} (\bibinfo {year} {1989})}\BibitemShut
  {NoStop}%
\bibitem [{\citenamefont {Teller}\ \emph {et~al.}(1995)\citenamefont {Teller},
  \citenamefont {Ishikawa},\ and\ \citenamefont {Wood}}]{ref28}%
  \BibitemOpen
  \bibfield  {author} {\bibinfo {author} {\bibfnamefont {E.}~\bibnamefont
  {Teller}}, \bibinfo {author} {\bibfnamefont {M.}~\bibnamefont {Ishikawa}}, \
  and\ \bibinfo {author} {\bibfnamefont {L.}~\bibnamefont {Wood}},\ }in\
  \href@noop {} {\emph {\bibinfo {booktitle} {Proceedings Frontiers in Physical
  Symposium, Joint American Physical Society and American Association of
  Physics Teachers Texas Meeting}}}\ (\bibinfo  {publisher} {Lubbock, Texas},\
  \bibinfo {year} {1995})\ \bibinfo {note} {preprint
  UCRL-JC-122708}\BibitemShut {NoStop}%
\bibitem [{\citenamefont {Rusov}\ \emph {et~al.}({\natexlab{c}})\citenamefont
  {Rusov}, \citenamefont {Litvinov}, \citenamefont {Mavrodiev}, \citenamefont
  {Linnik}, \citenamefont {Vaschenko}, \citenamefont {Zelentsova},
  \citenamefont {Beglaryan}, \citenamefont {Tarasov}, \citenamefont
  {Chernezhenko}, \citenamefont {Smolyar}, \citenamefont {Molchinikolov},\ and\
  \citenamefont {Merkotan}}]{ref29}%
  \BibitemOpen
  \bibfield  {author} {\bibinfo {author} {\bibfnamefont {V.~D.}\ \bibnamefont
  {Rusov}}, \bibinfo {author} {\bibfnamefont {D.~A.}\ \bibnamefont {Litvinov}},
  \bibinfo {author} {\bibfnamefont {S.~C.}\ \bibnamefont {Mavrodiev}}, \bibinfo
  {author} {\bibfnamefont {E.~P.}\ \bibnamefont {Linnik}}, \bibinfo {author}
  {\bibfnamefont {V.~N.}\ \bibnamefont {Vaschenko}}, \bibinfo {author}
  {\bibfnamefont {T.~N.}\ \bibnamefont {Zelentsova}}, \bibinfo {author}
  {\bibfnamefont {M.~E.}\ \bibnamefont {Beglaryan}}, \bibinfo {author}
  {\bibfnamefont {V.~A.}\ \bibnamefont {Tarasov}}, \bibinfo {author}
  {\bibfnamefont {S.~A.}\ \bibnamefont {Chernezhenko}}, \bibinfo {author}
  {\bibfnamefont {V.~P.}\ \bibnamefont {Smolyar}}, \bibinfo {author}
  {\bibfnamefont {P.~O.}\ \bibnamefont {Molchinikolov}}, \ and\ \bibinfo
  {author} {\bibfnamefont {K.~K.}\ \bibnamefont {Merkotan}},\ }\href@noop {}
  {\enquote {\bibinfo {title} {The kamland-experiment and soliton-like
  georeactor. part 2. fundemental geophysical consequences},}\ }
  ({\natexlab{c}}),\ \bibinfo {note} {in preparation}\BibitemShut {NoStop}%
\bibitem [{\citenamefont {Gilmore}(1985)}]{ref30}%
  \BibitemOpen
  \bibfield  {author} {\bibinfo {author} {\bibfnamefont {R.}~\bibnamefont
  {Gilmore}},\ }\href@noop {} {\emph {\bibinfo {title} {Catastrophe Theory for
  Scientists and Engineers}}}\ (\bibinfo  {publisher} {New York - Chichester -
  Brisbane - Toronto Wiley-Interscience Publication, John WileySons},\
  \bibinfo {year} {1985})\BibitemShut {NoStop}%
\bibitem [{\citenamefont {Passos}\ and\ \citenamefont {Lopes}(2011)}]{ref31}%
  \BibitemOpen
  \bibfield  {author} {\bibinfo {author} {\bibfnamefont {D.}~\bibnamefont
  {Passos}}\ and\ \bibinfo {author} {\bibfnamefont {I.}~\bibnamefont {Lopes}},\
  }\href@noop {} {\bibfield  {journal} {\bibinfo  {journal} {J. Atmos.
  Sol.-Terr. Phys.}\ }\textbf {\bibinfo {volume} {73}},\ \bibinfo {pages} {191}
  (\bibinfo {year} {2011})}\BibitemShut {NoStop}%
\bibitem [{\citenamefont {Mininni}\ \emph {et~al.}(2001)\citenamefont
  {Mininni}, \citenamefont {Gomez},\ and\ \citenamefont {Mindlin}}]{ref32}%
  \BibitemOpen
  \bibfield  {author} {\bibinfo {author} {\bibfnamefont {P.~D.}\ \bibnamefont
  {Mininni}}, \bibinfo {author} {\bibfnamefont {D.~O.}\ \bibnamefont {Gomez}},
  \ and\ \bibinfo {author} {\bibfnamefont {G.~B.}\ \bibnamefont {Mindlin}},\
  }\href@noop {} {\bibfield  {journal} {\bibinfo  {journal} {Sol. Phys.}\
  }\textbf {\bibinfo {volume} {201}},\ \bibinfo {pages} {203} (\bibinfo {year}
  {2001})}\BibitemShut {NoStop}%
\bibitem [{\citenamefont {Aaron}()}]{ref11}%
  \BibitemOpen
  \bibfield  {author} {\bibinfo {author} {\bibfnamefont {S.~C.}\ \bibnamefont
  {Aaron}},\ }\href@noop {} {\enquote {\bibinfo {title} {Experimental probes of
  axions},}\ }\bibinfo {note} {ArXiv:1009.4718}\BibitemShut {NoStop}%
\end{thebibliography}%

\end{document}